\documentclass[12pt,preprint]{aastex}

\shorttitle{flare ribbons approach} \shortauthors{Li et al.}

\begin{document}

\title{Flare Ribbons Approach Observed by the \emph{Interface Region
Imaging Spectrograph} and the \emph{Solar Dynamics Observatory}}

\author{Ting Li\altaffilmark{1,2}, Jun Zhang\altaffilmark{1,2} \& Yijun Hou\altaffilmark{1,2}}

\altaffiltext{1}{Key Laboratory of Solar Activity, National
Astronomical Observatories, Chinese Academy of Sciences, Beijing
100012, China; liting@nao.cas.cn} \altaffiltext{2}{University of
Chinese Academy of Sciences, Beijing 100049, China}

\begin{abstract}

We report flare ribbons approach (FRA) during a multiple-ribbon
M-class flare on 2015 November 4 in NOAA AR 12443, obtained by the
\emph{Interface Region Imaging Spectrograph} and the \emph{Solar
Dynamics Observatory}. The flare consisted of a pair of main ribbons
and two pairs of secondary ribbons. The two pairs of secondary
ribbons were formed later than the appearance of main ribbons, with
respective time delays of 15 and 19 minutes. The negative-polarity
main ribbon spread outward faster than the first secondary ribbon
with the same polarity in front of it, and thus the FRA was
generated. Just before their encounter, the main ribbon was
darkening drastically and its intensity decreased by about 70 \% in
2 minutes, implying the suppression of main-phase reconnection that
produced two main ribbons. The FRA caused the deflection of the main
ribbon to the direction of secondary ribbon with a deflection angle
of about $60^{\circ}$. Post-approach arcade was formed about 2
minutes later and the downflows were detected along the new arcade
with velocities of 35$-$40 km s$^{-1}$, indicative of the magnetic
restructuring during the process of FRA. We suggest that there are
three topological domains with footpoints outlined by the three
pairs of ribbons. Close proximity of these domains leads to
deflection of the ribbons in agreement with the magnetic field
topology.

\end{abstract}

\keywords{Sun: activity---Sun: flares---Sun: transition region---
Sun: UV radiation}

\section{Introduction}

Solar flares are among the most intense manifestations of energy
release on the Sun. The magnetic reconnection process in the corona
produces accelerated particles, which flow down along the
reconnected field lines and generate flare ribbons in the lower
solar atmosphere. The intersection of newly reconnected field lines
with the chromosphere and transition region corresponds to flare
ribbons observed in H$\alpha$ and ultraviolet (UV) wavelengths
(Forbes et al. 2006; Shibata \& Magara 2011). The shape, location
and dynamics of flare ribbons, as well as their relationship to
magnetic fields provide a significant amount of information about
the three-dimensional (3D) magnetic configuration involved in the
reconnection process (Gorbachev et al. 1988; Savcheva et al. 2015).

Flare ribbons are widely observed to spread systematically outwards
from the polarity inversion line (PIL), which is explained by the
ascending reconnection site (Fletcher et al. 2011; Schmieder et al.
2015). Moreover, flare ribbons also exhibit the apparent elongation
motion along the PIL with the maximum apparent speed comparable to
the local Alfv{\'e}n speed (Qiu 2009; Li \& Zhang 2009; Pontin et
al. 2016; Priest \& Longcope 2017). Sometimes unidirectional
brightening propagation of two flare ribbons parallel to the PIL is
associated with an asymmetric filament eruption, that is the
``zipper" motion (Tripathi et al. 2006; Liu et al. 2009). Different
from the zipping brightening motion, flare ribbons are also observed
to move in opposite directions parallel to the PIL (Su et al. 2007;
Li \& Zhang 2014). Qiu et al. (2010, 2017) suggested that a large
magnetic guide field parallel to the reconnection electric field may
be present during the elongation motion of flare ribbons. Priest \&
D{\'e}moulin (1995), Masson et al. (2009) and Aulanier et al. (2006)
proposed that 3D slipping magnetic reconnection at a null point,
separator or quasi-separatrix layers (QSLs) can account for the
elongation motion. Recently, the apparent slipping motions of flare
loops and ribbons during eruptive flares were investigated by Li \&
Zhang (2014, 2015) and Dud{\'{\i}}k et al. (2014, 2016). The
slippage exhibited a quasi-periodic pattern with a period of 3$-$6
minutes and its speed reached several tens of km s$^{-1}$.

The theoretical models usually assume a very simple magnetic field
structure and generate two approximately parallel ribbons (e.g.,
standard flare model in the review of Shibata \& Magara 2011;
sheared arcade model in Somov et al. 2002). However, eruptive flares
generally occur in complicated multipolar active regions (ARs) and
display a three or multiple ribbon structure (Liu et al. 2015; Bamba
et al. 2017). Wang et al. (2014) reported two successive eruptive
flares with three ribbons parallel to the PIL and suggested that
magnetic reconnection in a fan$-$spine magnetic topology produced
the complicated ribbons. Multiple-ribbon flares consists of two
parallel ribbons and several other secondary ribbons far from the
center of the AR (Mandrini et al. 2006; Zheng et al. 2015, 2016).
Zhang et al. (2014) found that more than a half of X-class flares in
the 24th solar cycle were associated with secondary ribbons and
suggested multiple sets of magnetic loops were involved in the
reconnection process (Chandra et al. 2009). Direct magnetic
connections between secondary ribbons and the main flare could be
responsible for the brightening of the secondary ribbons. These
connections are either by magnetic loops (Herant et al. 1989; Maia
et al. 2003) or by a system of null points, separatrices, and QSLs
at high altitude (Schrijver \& Title 2011). The brightness
enhancement of secondary ribbons is generally weaker than the main
central flare ribbons and appears later than the main ribbons (Zhang
et al. 2014; Schmieder et al. 2015).

The observations with high temporal and spatial resolutions from the
\emph{Interface Region Imaging Spectrograph} (\emph{IRIS}; De
Pontieu et al. 2014) allow an assessment of the low atmospheric
structure in unprecedented detail. In this paper, we report a
multiple-ribbon flare with GOES-class M3.7 in the flare productive
NOAA AR 12443 on 2015 November 4. Two pairs of secondary ribbons are
successively formed after the appearance of main ribbons and the
approach between one main ribbon and a secondary ribbon residing at
the same magnetic polarity is firstly presented. To our knowledge,
the observations of flare ribbons approach (FRA) have never been
reported before. We will investigate the dynamic evolution and
magnetic connectivity of multiple ribbons, and focus on the details
and physical process of the FRA.

\section{Observations and Data Analysis}

We combine data from the \emph{Solar Dynamics Observatory}
(\emph{SDO}; Pesnell et al. 2012) and the \emph{IRIS} to investigate
the multiple-ribbon flare and the process of FRA. The Atmospheric
Imaging Assembly (AIA; Lemen et al. 2012) on board the \emph{SDO}
provides multi-filter, near-simultaneous full-Sun images with a
resolution of $\sim$0$\arcsec$.6 per pixel and a cadence of 12
seconds. The observations of AIA 1600, 304, 171 and 131 {\AA}
channels are used in this work. The relations of flare ribbons with
magnetic fields are also studied by using the full-disk
line-of-sight (LOS) magnetograms from the Helioseismic and Magnetic
Imager (HMI; Scherrer et al. 2012). We concentrate on the
\emph{IRIS} slit-jaw images (SJIs) of 1400 {\AA} with a high spatial
sampling of $\sim$0$\arcsec$.33 per pixel and a cadence of about 12
seconds. The field-of-view (FOV) of the SJIs covers the central part
of the flaring region, and the process of FRA can be well detected.
The SJI 1400 {\AA} images are dominated by Si {\sc iv} line emission
from the transition region and continuum emission from the low
chromosphere (Tian et al. 2014). The \emph{IRIS} slit crosses the
newly formed arcade during the approach process in a large, coarse,
16-step raster covering a FOV of 30$\arcsec$$\times$119$\arcsec$ in
50 s with 2$\arcsec$ step size. Each raster step takes about 3 s,
and the spectral sampling is $\sim$0.05 {\AA} per pixel. The
\emph{IRIS} spectra of the Si {\sc iv} $\lambda$1402.77 line were
analyzed and fitted by a single-Gaussian function to obtain the
Doppler shifts along the arcade (Peter et al. 2014).

\section{Results}

\subsection{Flux Rope Eruptions and Evolution of Multiple Ribbons}

The analyzed M3.7 flare on 2015 November 4 was initiated at 13:31 UT
and peaked at 13:52 UT from the GOES SXR 1$-$8 {\AA} flux data
(Figure 4(a)). About 10 minutes before the flare started, a filament
at the north of AR 12443 was activated and showed an evident
clockwise rotation around the filament axis (Figure 1(a); see
Animation 304-eruption). The filament gradually developed into two
filament structures F1 and F2 as seen from the 304 {\AA} images
(Figure 1(b)). Then the filament materials of F1 and F2 moved in
opposite directions from their crossing location (white arrows in
Figure 1(b)), implying the occurrence of tether-cutting reconnection
(Moore et al. 2001; Chen et al. 2016). We obtained the stack plots
along the moving directions of filament materials and then
calculated the bidirectional velocities of about 150 km s$^{-1}$ and
210 km s$^{-1}$ for F1 and F2. Meanwhile, two quasi-parallel flare
ribbons R1 and R2 appeared underlying the eruptive filament F1.
Starting from 13:32 UT, R2 at the east appeared and propagated to
southeast at a speed of about 120 km s$^{-1}$ (green arrows in
Figures 1 (c)-(d)), which finally formed a hook-shaped segment with
a length of about 80 Mm. In the high-temperature channel of 131
{\AA} (about 11 MK), two flux ropes FR1 and FR2 were detected
(Figures 1 (e)-(f); see Animation 131-eruption), with their lower
parts corresponding to the two filaments F1 and F2 (Li \& Zhang
2013; Cheng et al. 2014). The flux rope FR1 exhibited an asymmetric
eruption with its eastern footpoints slipping along the hook of R2
(Figures 1 (g)-(h)) and finally delineated a ``triangle-shaped flag"
surface. The observations are similar to the event presented by Li
\& Zhang (2014) and imply the occurrence of slipping magnetic
reconnection. R1 resided at negative-polarity magnetic fields and R2
at positive-polarity fields (Figure 1(e)).

About 15 minutes after the flare onset ($\sim$ 13:46 UT), a pair of
secondary ribbons PSR1 and NSR1 appeared adjacent to two main
ribbons R1 and R2 (Figures 2(a)-(b)). Later on, the interaction
between ribbons R1 and NSR1 was initiated when R1 chased up NSR1
(Figure 2(c)). The detailed analysis of the approach process is
carried out in Figures 3-5. The west cusp of ribbon R2 extended to
the north and encountered the ribbon PSR1 (green arrows in Figure
2(c)-(d)). Then R2 and PSR1 approached each other and a brightening
between the R2 and PSR1 ribbons was observed (see 13:50 UT in
Animation 304-eruption). Afterwards the brightening propagated to
the north along the hook-shaped segment of R2. Another pair of
secondary ribbons PSR2 and NSR2 were detected about 4 minutes after
the appearance of PSR1 and NSR1 (Figure 2(c)). Afterwards, the
intensities and widths of PSR2 and NSR2 were both increased at 13:56
UT (Figure 2(d). The 131 {\AA} observations showed the existence of
hot coronal loops connecting PSR1 and NSR1 (Figures 2(f)-(g)),
indicative of their conjugated property. Similarly, the
high-temperature loops connecting PSR2 and NSR2 can also be detected
(red dotted curves in Figure 2(h)).

The overlay of the flare ribbon time evolution over the magnetogram
is displayed in Figure 2(e). The color indicates the time of the
ribbon brightness observed at \emph{SDO} 1600 {\AA} and \emph{IRIS}
1400 {\AA}. The west ribbons NSR1 and NSR2 anchored at the
negative-polarity magnetic fields and the east ones at the
positive-polarity fields. It shows that the generally parallel
portions of two main ribbons R1 and R2 exhibit an evident separation
motion with the high-temperature flare loops connecting them. Two
pairs of secondary ribbons are formed later than the main ribbons.
Each pair of secondary ribbons has corresponding coronal loops that
appear right after the formation of these ribbons. We suggest that
there are three topological domains with footpoints outlined by the
three pairs of ribbons. The approach process of ribbons R1 and NSR1
is interpreted as the interaction of two different topological
domains.

\subsection{Approach Process Between Two Ribbons}

The ribbons R1, NSR1 and NSR2 at the west were clearly observed by
the \emph{IRIS}, and thus the details of flare ribbon interactions
can be well obtained. The main ribbon R1 continually spread towards
the southwest as the flare developed (Figures 3(a)-(c); see
Animation 1400-approach). At 13:46 UT, the secondary ribbon NSR1 was
formed in front of R1 and was composed of numerous slipping dot-like
substructures (Figure 3(b)). Later on, the arch-shaped ribbon NSR1
moved southward with a smaller speed than R1, resulting in the
distance between NSR1 and R1 becoming smaller and smaller. About 4
minutes later, another secondary ribbon NSR2 appeared at the south
of NSR1 (Figure 3(c)). Then the three ribbons R1, NSR1 and NSR2 kept
moving to the south. Starting from 13:51 UT, the ribbon R1 was
darkening with the intensity decrease of about 70 \% when R1 moved
close to NSR1 (Figure 3(d)). From then on, the ribbon R1 continually
approached to NSR1 and evident brightening appeared at the merging
location (Figure 3(e)). From about 13:55 UT, the brightening at the
west was detected to propagate from the interaction location to the
west (black arrows in Figures 3(f)-(h)). The direction of
brightening propagation was almost consistent with NSR1 (Figure
3(i)). After examining the 1400 {\AA} data, we suggest that the
newly formed brightening could be associated with the evolving
ribbon R1 deflected by the NSR1 ribbon by about $60^{\circ}$ (see
Figure 3(i)).

The spatio-temporal relations among the three ribbons R1, NSR1 and
NSR2 were presented in Figure 4. The propagating velocity of R1
decreased from 13.1 km s$^{-1}$ at 13:38 UT to 4.8 km s$^{-1}$ at
13:51 UT just before the approach (Figure 4(b)). About 15 minutes
after the flare started, NSR1 appeared and propagated southward at a
speed of about 1.5 km s$^{-1}$ (Figure 4(a)). NSR2 was formed about
19 minutes after the flare onset and moved in front of NSR1 at a
velocity of 4.3 km s$^{-1}$. Then evident darkening process of R1
was started at about 13:51 UT when R1 and NSR1 were very close to
each other (distance of 1.2 Mm) and the intensity decreased by 70 \%
in 2 minutes (Figure 4(b)). The FRA of R1 and NSR1 were initiated at
13:53 UT (blue dash-dotted lines in Figure 4) and caused the
brightening at the interaction location. Afterwards, the ribbon R1
was deflected to the direction of NSR1. As seen from the stack plot
(Figure 4(c)) along slice ``C$-$D" (blue line in Figure 3(f)), there
were two times of propagations of deflected R1 with their velocities
reaching about 71 and 54 km s$^{-1}$.

Associated with the FRA, part of the arcade was newly formed at the
northeast of the interaction location from about 13:55 UT (Figures
5(a) and (d)). Then the mass flows from the top of the arcade to its
two ends appeared and a large part of the arcade could be detected
in 171 {\AA} and 1400 {\AA} channels (Figures 5(b) and (e)). The
east end of the arcade was located at the positive-polarity fields
and the west end anchored in the negative-polarity fields (contours
in Figure 5(b)). The top part of the arcade could also be observed
at 1600 {\AA} (Figure 5(c)), implying the lower-temperature
component of the arcade. At about 13:59 UT, the arcade became more
extended and its west leg was connected with the deflected R1
(Figure 5(f)). By plotting the space-time map of 1400 {\AA} (Figure
5(g)) along slice ``E$-$F" (Figure 5(f)), we have inferred the
speeds of downflows with values of 35 and 40 km s$^{-1}$. As seen
from the space-time map, the arcade was formed about 2 minutes later
than the initiation of FRA (blue and white dash dotted lines), and
thus we name the newly formed arcade as post-approach arcade (PAA).
The spectra of Si {\sc iv} $\lambda$1402.77 line at three different
locations (diamonds in Figures 5(e)-(f)) of PAA were displayed in
Figures 5(h)-(i). At 13:57 UT, two east locations showed blueshifts
of about 9 and 12 km s$^{-1}$ while the west location exhibited a
redshift of 5 km s$^{-1}$ (Figure 5(h)). About 2 minutes later, the
blueshifts at two east locations reached about 13 and 11 km s$^{-1}$
and the redshift at the west leg increased to 8 km s$^{-1}$ (Figure
5(i)).

\section{Summary and Discussion}

We present the \emph{IRIS} and \emph{SDO} observations of the
approach process between one main ribbon R1 and a secondary ribbon
NSR1 with the same magnetic polarity during a multiple-ribbon M3.7
flare on 2015 November 4. The two main ribbons of the flare had a
portion generally parallel to each other and also a hook-shaped
segment at the east of the positive-polarity ribbon, which
propagated towards the southeast at a speed of about 120 km
s$^{-1}$. Meanwhile the east footpoints of the associated flux rope
displayed a slipping motion along the hook, delineating a
``triangle-shaped flag surface". Two pairs of secondary ribbons were
formed successively at the periphery of two main ribbons in the late
stage of the impulsive phase. R1 spread outward much faster than
NSR1 in front of it and the distance between the two ribbons became
smaller and smaller. R1 began to darken evidently and its intensity
decreased by 70 \% in 2 minutes when it came very close to NSR1
(less than 1.2 Mm). Then R1 approached to NSR1 and caused the
deflection of ribbon R1 with a deflection angle of about
$60^{\circ}$. The deflected R1 propagated in the direction of NSR1
at velocities of 50$-$70 km s$^{-1}$. A new loop structure PAA was
formed during the FRA, with its west leg connecting with the
deflected R1. The downflows along the PAA had Doppler shifts of
about 10 km s$^{-1}$ and projected velocities of 40 km s$^{-1}$ in
the plan of sky.

The secondary ribbons were formed in pairs and sequentially at the
periphery of main ribbons after the flare onset, with respective
time delays of 15 and 19 minutes for the first and second pairs of
secondary ribbons. The conjugate footpoints of each pair of
secondary ribbons were connected by high-temperature coronal loops.
Previous studies suggested that the initial magnetic reconnections
in main flare regions triggered the high-latitude null-point
reconnections in a fan-spine-like topology (Joshi et al. 2015; Liu
et al. 2015) or long-distance loop-loop interactions (Maia et al.
2003; Chandra et al. 2009) and thus produced the secondary ribbons.
The secondary ribbons in our observations were composed of numerous
small-scale slipping substructures, which correspond to the
footpoints of reconnected field lines (Li \& Zhang 2015). We suggest
that the flux rope eruption pushes overlying loops against the
neighboring topological domain and then magnetic reconnections
between the erupting domain and its neighboring one result in the
formation of the first pair of secondary ribbons. Later on, another
topological domain is disturbed after the appearance of ribbons PSR1
and NSR1 and generates the second pair of secondary ribbons.

Associated with the FRA of ribbons R1 and NSR1, several new
observational features have been identified. The ribbons R1 and NSR1
corresponded to the ends of reconnected magnetic lines in two
different topological domains (Fletcher et al. 2011; Schmieder et
al. 2015). Associated with the separation motion of two main
ribbons, magnetic reconnection in the center region was gradually
spreading towards the border of the topological domain. When
magnetic reconnection occurred at the border of the topological
domain, it encountered with another neighboring topological domain
and resulted in the phenomenon of FRA. We suggest that the
interaction location corresponded to the boundary of two different
domains. The close proximity of these domains caused the deflection
of ribbon R1 with a deflection angle of about $60^{\circ}$. The
deflected ribbon propagated in the direction consistent with NSR1,
in agreement with the magnetic topology of the neighboring domain.
The rapid ``darkening" of R1 just before the approach implied that
coronal magnetic reconnection was suppressed at the boundary of
different magnetic systems. Furthermore, the newly formed loop
structure PAA indicated that magnetic restructuring was involved in
the process of FRA. The downflows along the PAA were simultaneously
detected based on the imaging and spectral observations, which
probably corresponded to reconnection downflows. However, the
triggering mechanism of secondary ribbons and physical process of
FRA are still open questions. In the future, more observational and
simulation studies are required to understand the process of FRA as
well as other complicated dynamic evolution of solar flares.

\acknowledgments {We are grateful to Kai Yang and Shu-Hong Yang for
useful discussions. \emph{SDO} is a mission of NASA's Living With a
Star Program. \emph{IRIS} is a NASA small explorer mission developed
and operated by LMSAL with mission operations executed at NASA's
Ames Research center and major contributions to downlink
communications funded by the Norwegian Space Center (NSC, Norway)
through an ESA PRODEX contract. This work is supported by the
National Natural Science Foundations of China (11773039, 11533008,
11673035 and 11673034) and the Youth Innovation Promotion
Association of CAS (2017078).}

{}
\clearpage

\begin{figure}
\centering
\includegraphics
[bb=8 120 559 703,clip,angle=0,scale=0.8]{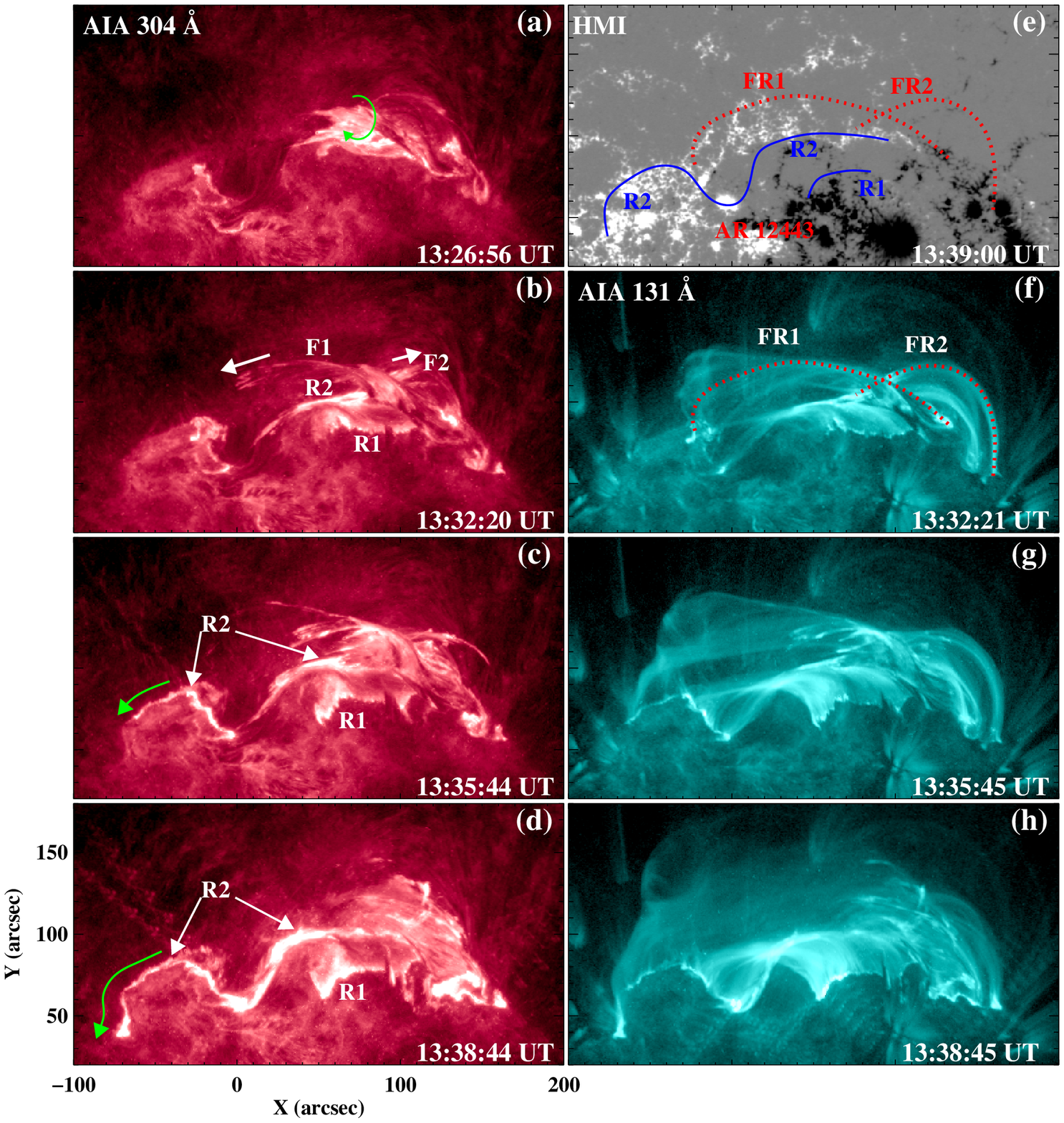} \caption{Time
sequences of \emph{SDO}/AIA 304 {\AA} images, 131 {\AA} images and
\emph{SDO}/HMI LOS magnetogram showing the evolution of flux ropes
and the M3.7 flare in AR 12443 on 2015 November 4 (see Animations
304-eruption and 131-eruption). The green arrow in panel (a) denotes
the clockwise rotation of the filament and the white arrows in panel
(b) represent the bidirectional flows along two filaments F1 and F2.
R1 and R2 are two main flare ribbons, and green arrows in panels
(c)-(d) point to the propagating direction of the hooked segment of
R2. Red dotted lines in panels (e)-(f) outline the main axes of flux
ropes FR1 and FR2. Blue solid lines in panel (e) show the
morphologies of ribbons R1 and R2 at 13:39 UT. \label{fig1}}
\end{figure}
\clearpage

\begin{figure}
\centering
\includegraphics
[bb=26 140 570 745,clip,angle=0,scale=0.8]{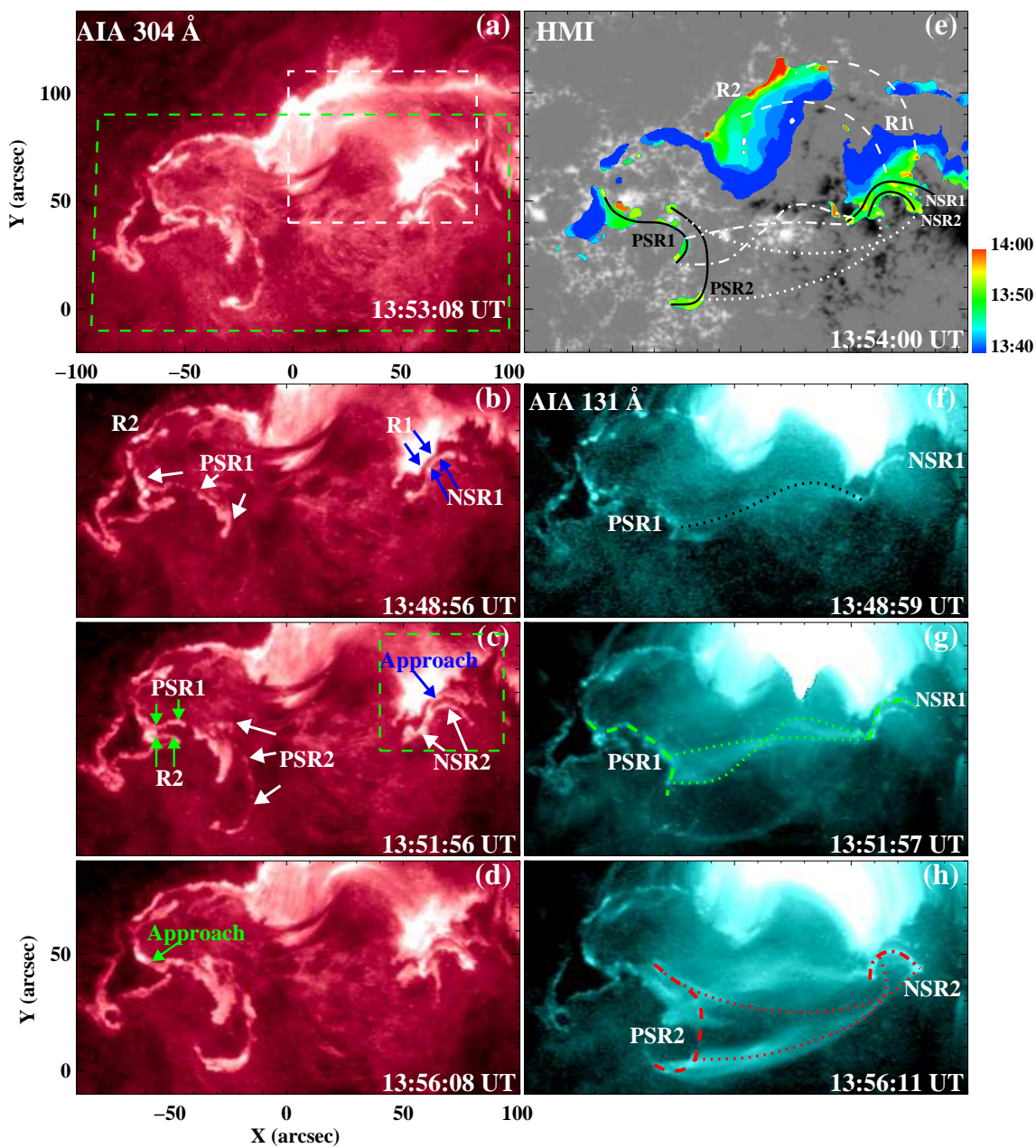}
\caption{Appearance of two pairs of secondary ribbons (PSR1-NSR1,
PSR2-NSR2) and their interactions with main ribbons R1 and R2 in the
late phase of the flare. The green and white rectangles in panel (a)
respectively denote the FOVs of the images in panels (b)-(d), panels
(f)-(h) and Figure 5. The green rectangle in panel (c) shows the FOV
of Figure 3. Panel (e) shows the \emph{SDO}/HMI LOS magnetogram with
an overlay of the positions of newly brightened ribbons. The color
indicates the time of the ribbon brightness from 13:40 UT to 14:00
UT in the images of 1600 {\AA} and 1400 {\AA}. Green lines in panel
(g) show the first pair of secondary ribbons PSR1 and NSR1 and
high-temperature loops connecting them. Red lines in panel (h)
denote the second pair of secondary ribbons PSR2 and NSR2 and their
connecting loops. Their duplications are also shown in panel (e).
\label{fig2}}
\end{figure}
\clearpage

\begin{figure}
\centering
\includegraphics
[bb=22 163 546 653,clip,angle=0,scale=0.8]{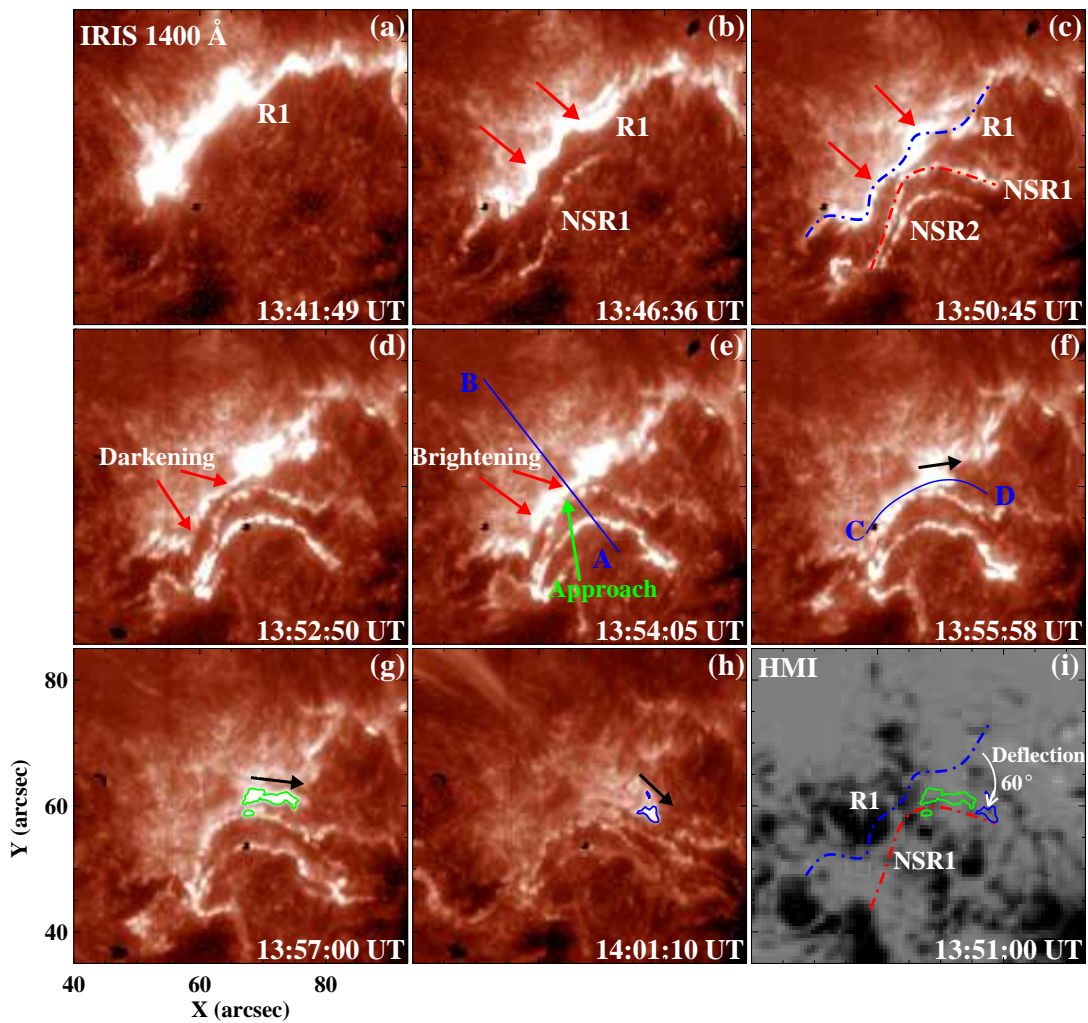} \caption{Time
sequence of \emph{IRIS} 1400 {\AA} images and \emph{SDO}/HMI LOS
magnetogram showing the approach process between ribbons R1 and NSR1
(see Animation 1400-approach). Red arrows in panels (b)-(c)
represent the propagating direction of R1. Blue and red dash-dotted
lines in panel (c) outline ribbons R1 and NSR1, and are duplicated
in the magnetogram of panel (i). Blue lines ``A$-$B" and ``C$-$D"
show the cut positions used to obtain the time-distance plots shown
in Figures 4(a) and (c). Black arrows in panels (f)-(h) denote the
propagation of deflected ribbon after the ribbons approach (FRA) of
R1 and NSR1. The green and blue curves in panels (g)-(h) are the
brightness of deflected ribbon, and their duplications are also
shown in panel (i). \label{fig3}}
\end{figure}
\clearpage

\begin{figure}
\centering
\includegraphics
[bb=89 109 512 714,clip,angle=0,scale=0.85]{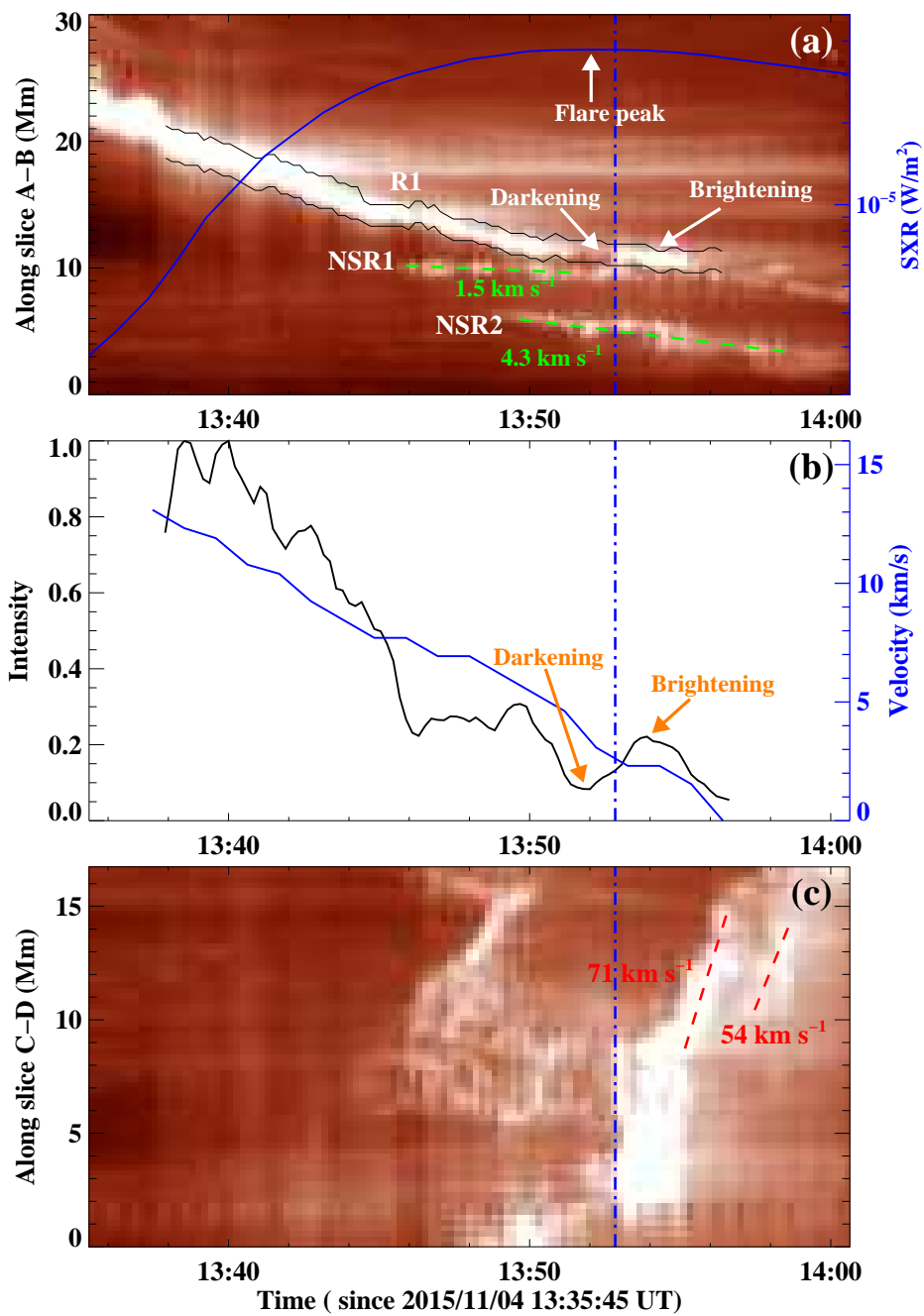} \caption{Panel
(a): time-distance plot along slice ``A$-$B" (blue line in Figure
3(e)) at 1400 {\AA}. The blue curve shows GOES SXR 1$-$8 {\AA} flux
of the M3.7 flare. Blue dash-dotted line denotes the start time of
FRA between R1 and NSR1. Panel (b): evolution of intensity (black
curve) and propagating velocity (blue curve) of R1. The calculated
region of the intensity and velocity is within the black curves in
panel (a). Panel (c): time-distance plot along slice ``C$-$D" (blue
line in Figure 3(f)). Red dashed lines outline the propagations of
the deflected ribbon. \label{fig4}}
\end{figure}
\clearpage

\begin{figure}
\centering
\includegraphics
[bb=28 167 549 625,clip,angle=0,scale=0.9]{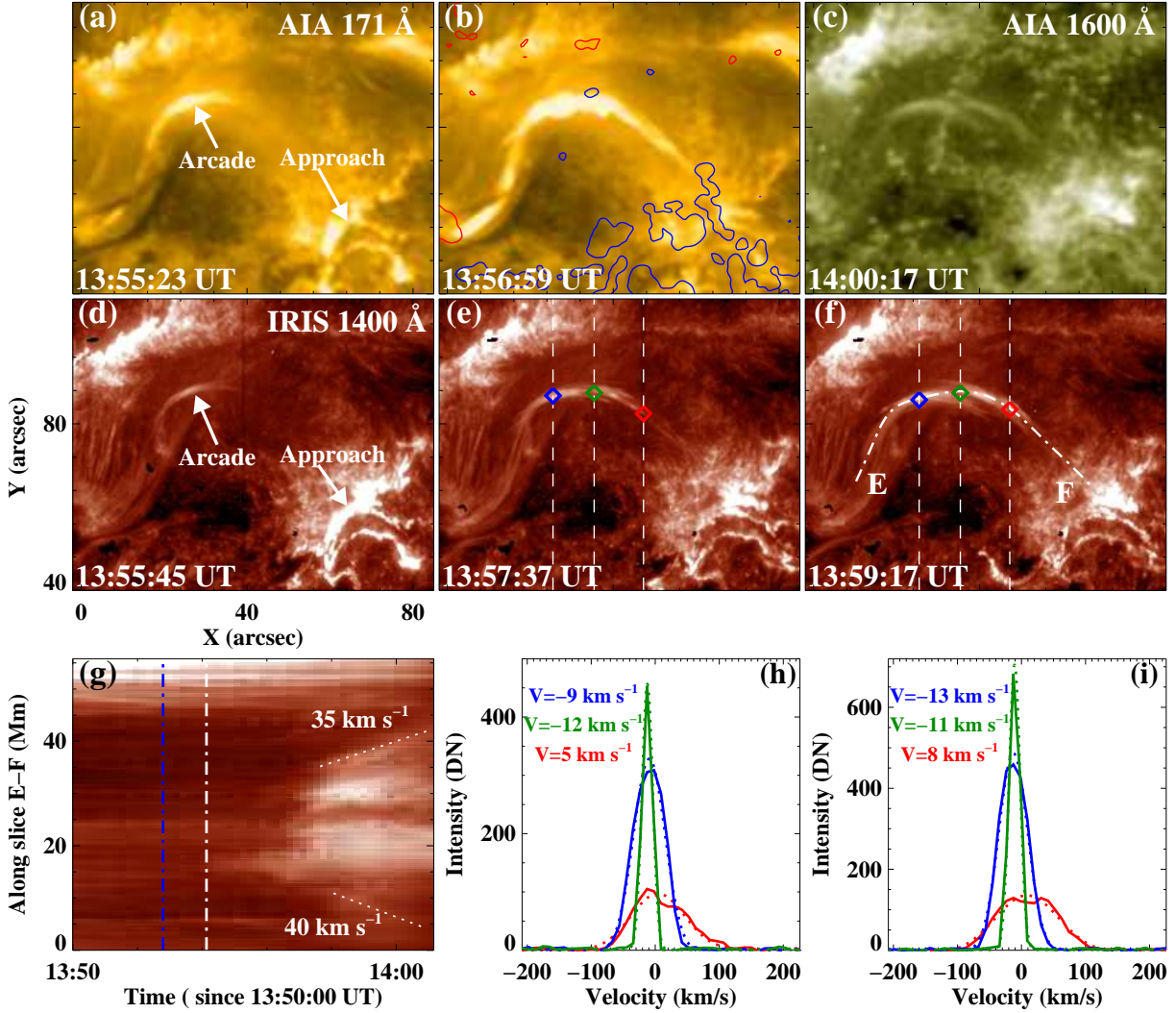} \caption{Panels
(a)-(f): multi-wavelength observations of the post-approach arcade
(PAA) during the approach process. The red and blue contours in
panel (b) are the magnetic fields at $\pm$500 G levels at the ends
of the PAA. Dashed lines in panels (e)-(f) show the slit locations
and the diamonds denote the intersections of the PAA and the slit.
Dash-dotted curve ``E$-$F" in panel (f) shows the cut position used
to obtain the time-distance plot shown in panel (g). Panel (g):
time-distance plot along slice ``E$-$F" at 1400 {\AA}. Blue and
white dash dotted lines respectively represent the start of FRA and
the appearance of PAA. Panel (h): profiles of the Si {\sc iv}
$\lambda$1402.77 line at different locations (diamonds in panel (e))
along the PAA. The dotted curves are the corresponding
single-Gaussian fitting profiles. Panel (i): similar to panel (h)
but for the diamond locations in panel (f). \label{fig4}}
\end{figure}
\clearpage

\end{document}